\documentclass[%
 reprint,
 amsmath,amssymb,
 aps,
]{revtex4-1}

\usepackage{mathrsfs}
\usepackage{color}
\definecolor{mygreen}{rgb}{0, 0.6, 0}

\def\bea{\begin{eqnarray}}
\def\eea{\end{eqnarray}}

\usepackage{graphicx}
\usepackage{dcolumn}
\usepackage{bm}

\begin{document}


\title{Cellular wrapping of elastic particles by a supported lipid membrane}

\author{Amir Khosravanizadeh}
\email{amir.khosravanizadeh@ijm.fr}
\affiliation{Université Paris Cité, CNRS, Institut Jacques Monod, F-75013 Paris, France}
\author{Pierre Sens}
\affiliation{Institut Curie, Université PSL, Sorbonne Université, CNRS UMR168, Physique des Cellules et Cancer, 75005 Paris, France}
\author{Farshid Mohammad-Rafiee}
\email{farshid@iasbs.ac.ir}
\affiliation{Department of Physics, Institute for Advanced Studies in Basic Sciences (IASBS), Zanjan 45137-66731, Iran}
\affiliation{Department of Physics and Astronomy, University of Pennsylvania, Philadelphia, Pennsylvania, USA
}

\date{\today}

\begin{abstract}
Constancy of life vitally depends on the internalization of particles through biomembranes. Of particular interest, cellular uptake, including phagocytosis, receptor-mediated endocytosis, and membrane fusion, critically depends on the elasticity of particles. Cellular membranes are strongly linked to a supporting cytoskeleton. However, in most previous studies, the effect of this cortical network somehow is overlooked. In this paper, we study the cellular wrapping of a membrane around a 2D elastic particle in the presence of a substrate mimicking cytoskeleton. Our simulations show that the impact of particle flexibility on the wrapping process depends on the magnitude of the membrane–particle adhesion. In contrast, the extent of membrane protrusions formed around the target always increases with target stiffness. Since the extension of membrane protrusions is an essential step in the phagocytosis process, this result may indicate a selective behavior of macrophages in the phagocytosis of aged red blood cells.
\end{abstract}

\pacs{Valid PACS appear here}
\maketitle
Cells, as the fundamental units of life, are separated from their environment by a self-assembled structure of lipid molecules named membrane. They can uptake external particles through the lipid membrane with different mechanisms\cite{lodish}. Cellular internalization encompasses a range of distinct pathways, from the large-scale, actin-driven processes of phagocytosis and macropinocytosis to endocytic routes such as clathrin-mediated and caveolar-mediated endocytosis. While the former are characterized by active, actin-mediated membrane remodeling, in the latter, the cytoskeleton plays a less prominent role and the processes are often modeled as passive processes\cite{conner2003regulated}. Basically, in most of these mechanisms, the adhesion interaction of particle's ligands with diffusive membrane's receptors must balance by the elastic energy of the membrane, which includes bending and tension\cite{yi2017kinetics,huang2013role}. Previous studies have shown that the engulfment of the membrane around a target depends on the size\cite{chithrani2006determining,zhao2011interaction,hashemi2014regulation,
deserno2004elastic,bahrami2014wrapping,zhang2009size}, shape\cite{chithrani2006determining,gao2005mechanics,
vacha2011receptor,chen2016shape,bahrami2014wrapping,bahrami2013orientational,dasgupta2014membrane}, and surface properties of the particle\cite{zhao2011interaction,ding2012role}. Consequently, these results led to biomedical applications of nanoparticles in different fields, such as chemotherapy, bioimaging, biosensing, and drug and gene delivery\cite{xia2008nanomaterials,weissleder2006molecular,nel2009understanding,allen2004drug,whitehead2009knocking,peer2007nanocarriers}.

Stiff and rigid particles have been investigated in a large number of studies, and it is acceptable in the case of some industrial nanoparticles\cite{hashemi2014regulation,deserno2004elastic,vacha2011receptor,chen2016shape,bahrami2014wrapping,bahrami2013orientational,dasgupta2014membrane}. However, most biological functions include interactions of lipid membranes with soft targets, vesicles, and micelles\cite{lodish}. It has been found that soft particles, such as young red blood cells\cite{marik1993effect,xu2018stiffness}, cannot be phagocytosed by macrophages, whereas stiffer ones can easily be digested\cite{beningo2002fc,tao2005micromachined}. On the other hand, there is convincing evidence that the rigidity of HIV viruses capsid is regulated during budding out or entering into a host\cite{kol2006mechanical,kol2007stiffness}. Furthermore, an effective strategy to improve the efficiency of drug delivery capsules could be controlling their stiffness\cite{geng2007shape,masoud2012controlled}. Therefore, some studies were performed to probe the connection between the engulfment of vesicles and their flexibility, but the results are unclear and sometimes inconsistent\cite{beningo2002fc,tao2005micromachined,banquy2009effect,liu2012uptake,yi2011cellular,yue2013molecular,yi2017kinetics}.\\
\begin{figure}[b!]
\centering
\includegraphics[width=1\columnwidth]{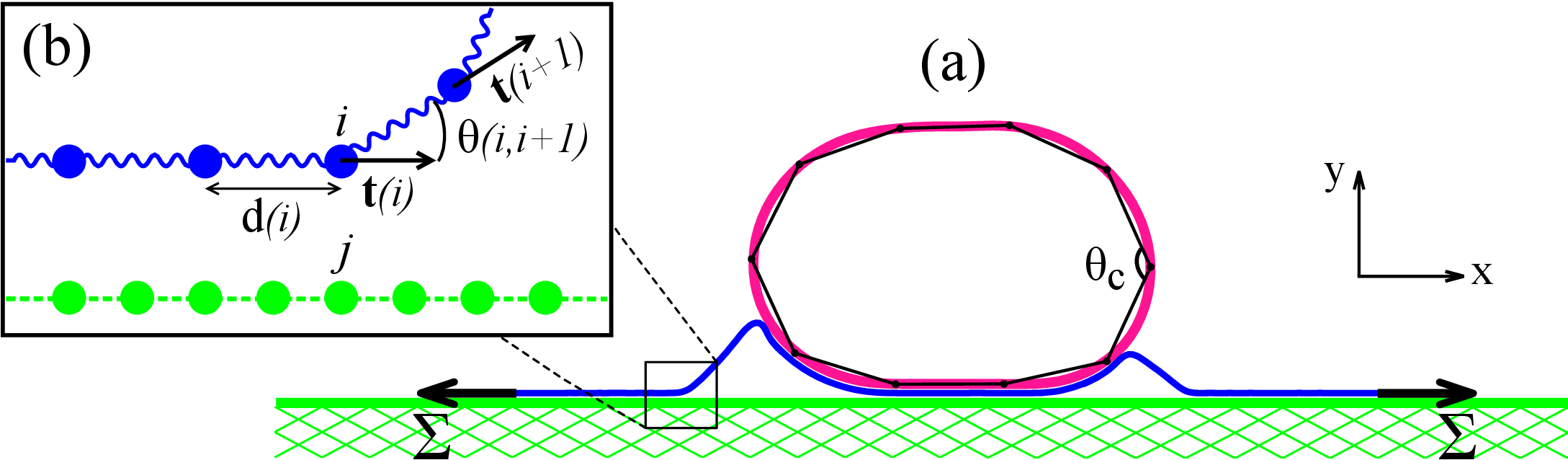}
\caption{(a) The schematic picture of a 2D flexible vesicle (pink) engulfed by a flat membrane (blue) attached to a non-deformable cytoskeleton substrate (green). The black network in the vesicle represents an inner meshwork that controls the elasticity of the vesicle. The total length of the particle is fixed and it's flexibility modulates with parameter $\kappa_c$. (b) a closed view of the bead-spring membrane model.}
\label{fig1}
\end{figure}
\begin{figure*}[ht!]
\centering
\includegraphics[width=1.5\columnwidth]{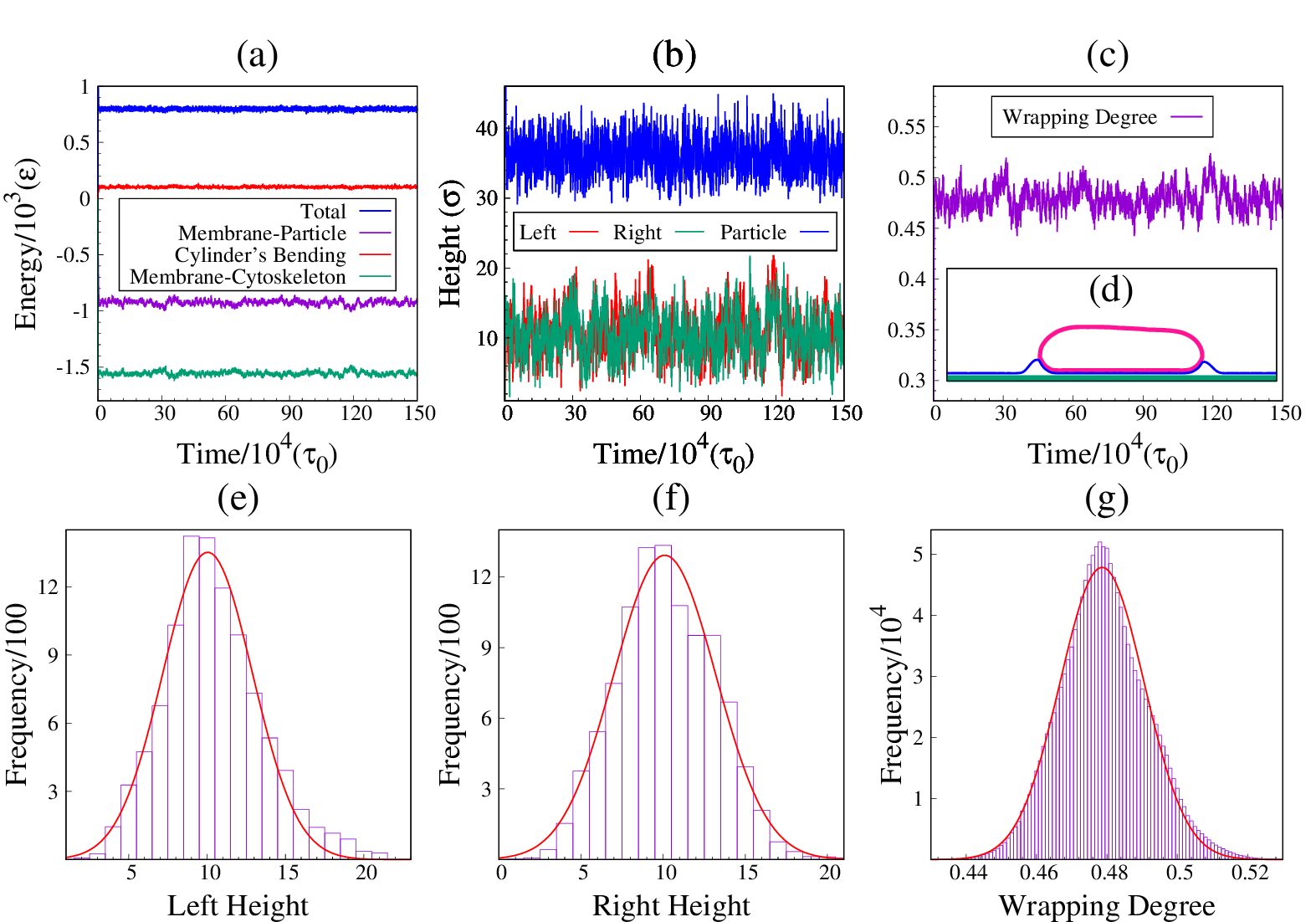}
\caption{The simulation results for the membrane-particle adhesion energy $\omega=6.14\, \varepsilon/{\sigma}^2$ and particle stiffness $\kappa_c=200\, \varepsilon$. Time evolution of (a) the total energy of simulation (blue), membrane-particle adhesion energy (purple), membrane-cytoskeleton adhesion energy (green), and bending energy of the particle (red); (b) the height of the membrane protrusions in the left (red) and right (green) sides of the particle, and the maximum height of the particle with respect to the cytoskeleton (blue); (c) the wrapping degree of the target. Panel (d) shows a typical snapshot of the membrane conformation in the equilibrium state. Panels (e) and (f) correspond to the distribution of the membrane's height in the left and right side of the particle, respectively. The distribution of the equilibrated wrapping degree is displayed in the panel (g).}
\label{fig2}
\end{figure*}
Experimentally, some cases show a higher wrapping rate for stiffer particles than softer ones\cite{beningo2002fc,tao2005micromachined}. In contrast, another study found the opposite behavior\cite{liu2012uptake}. Furthermore, some experiments have shown that nanoparticles with intermediate flexibility are engulfed more rapidly than either softer or stiffer particles\cite{banquy2009effect}.\\
Theoretically, Yi et al. demonstrated that the engulfment of elastic nanoparticles requires more adhesion energy than stiffer ones\cite{yi2011cellular}. They have also investigated the interaction of soft particles with a lipid vesicle which could lead to different wrapping phases\cite{yi2016incorporation}. Moreover, a molecular dynamics simulation by Yue and Zhang determined that softer particles can be wrapped faster than stiffer ones. They also found several different pathways for entering soft vesicles into cells, which depend on membrane-vesicle adhesion\cite{yue2013molecular}.\\
In most previous studies, the wrapping process has typically been considered as the interaction between a vesicle and a free membrane that is not attached to anything else\cite{deserno2004elastic,yi2011cellular,yue2013molecular}. However, it is known that a highly essential part of living cells is their cortical cytoskeleton. The cytoskeleton emplaces underneath the lipid membrane, and one of its vital roles is restricting the membrane to shape the cell\cite{lodish}. Furthermore, \textit{in vitro} studies of membrane-particle interactions often use supported bilayers attached to a substrate.\\
In the previous paper, we introduced a coarse-grained model of the membrane, which is appropriate for large-scale simulations in constant surface tension\cite{khosravani}. Using this model, we studied the wrapping process of a supported membrane around a long rigid cylindrical particle. The supporting substrate underneath the lipid membrane mimics the effect of a passive cytoskeleton. Then, by changing the circular cross-section of the cylinder to an elliptical one, we studied the role of particle shape and orientation in the engulfment process. We found that the membrane wrapping proceeds differently whether the initial contact occurs at the target’s highly curved part or along its long side\cite{khosravani2}. The present study investigates the effect of target flexibility in the same system. (see Fig. \ref{fig1})\\
Briefly, our model is obtained from the discretization of the Helfrich energy\cite{helfrich1973elastic} in 2D, which can be simulated as a collection of beads and springs using MD and MC methods(see Supporting Information). It is tunable with two elastic parameters: 1) the bending rigidity of the membrane, $\kappa$, and  2) its surface tension, $\Sigma$. The target and the cytoskeleton are also constructed as collections of beads. The cytoskeleton is fixed in space while the target is free to move in the $x-y$ plane. The total length of the particle is fixed, and its flexibility modulates with parameter $\kappa_c$. An attraction potential between the beads is implemented to model the membrane-particle and membrane-cytoskeleton ligand-receptor interactions. The membrane adhesion energy with the target is characterized by the parameter $\omega$ and with the substrate by $\omega_s$. Here, the value of bending rigidity, $\kappa=20 \, \varepsilon$, and surface tension, $\Sigma=1.5 \, \varepsilon/\sigma^2$ are fixed, where $\sigma$ and $\varepsilon$ are the unit length and energy of the MD simulations. Then in the constant value of membrane-cytoskeleton adhesion, $\omega_s=1.5 \, \varepsilon/\sigma^2$, the wrapping process is investigated as a function of target flexibility, $\kappa_c$, for different values of target adhesion energy, $\omega$.\\ 
After some sufficient MD steps, the system reaches an equilibrium state where the representing parameters of the system fluctuate around their equilibrium values. The equilibrated state of the system can be specified by looking at the time variation of the energies, the wrapped length of the particle, and the height of the membrane protrusions wrapped around the target. The results of a typical simulation for $\kappa_c=200\, \varepsilon$ and $\omega=6.14\, \varepsilon/{\sigma}^2$ are represented in Fig. \ref{fig2}. Panel (a) shows the time course of the total system's energy (in blue), the membrane-particle adhesion energy (purple), the membrane-cytoskeleton adhesion energy (green), and the bending energy of the particle (red). Panel (b) represents the height of the membrane protrusions on the left (red) and right (green) sides of the target. The blue line indicates the maximum height of the target relative to the cytoskeleton as a reference point. Panel (c) shows the wrapping degree of the target (the wrapped length of the target, compared to its total length) as a function of time. A typical snapshot of the membrane conformation is displayed in panel (d). Panels (e) and (f) represent the membrane's height distributions on the left and right sides of the target, respectively. The distribution of the wrapping degree is also indicated in panel (g). These distributions are obtained from the equilibrated state of the system. Corresponding to panels (e) to (g), these distributions are Gaussian, which are in agreement with the thermal fluctuation of energy, heights, and wrapping degree in panels (a) to (c).\\ 
Fig. \ref{d_wrap} shows the behavior of the wrapping degree as a function of the particle stiffness, $\kappa_c$, for different values of target adhesion $\omega$.
\begin{figure}[ht!]
\centering
\includegraphics[width=1\columnwidth]{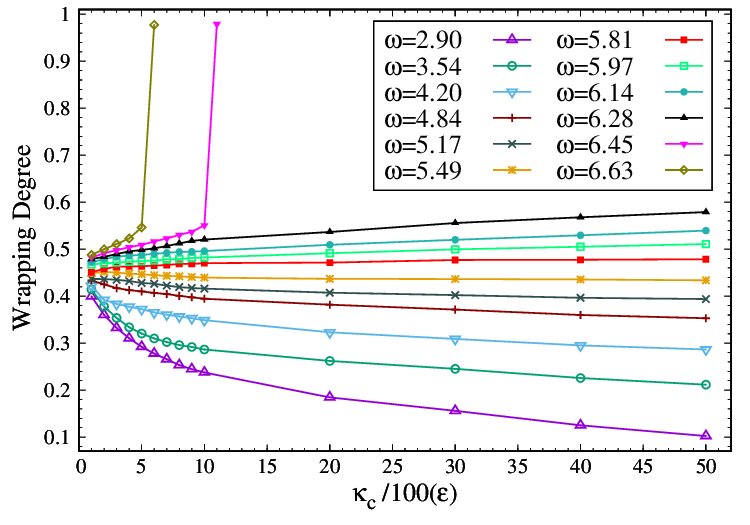}
\caption{The variation of the wrapping degree as a function of $\kappa_c$ for different values $\omega$.}
\label{d_wrap}
\end{figure}
This plot represents two different situations. If $\omega \lesssim 5.8 \, \varepsilon/{\sigma}^2$, increasing the target rigidity decrease the wrapping degree, while for $\omega \gtrsim 5.8 \, \varepsilon/{\sigma}^2$ the wrapping degree always is an increasing function of $\kappa_c$. This result can provide an explanation for different uptake rates of soft and rigid particles reported in the literature\cite{beningo2002fc,tao2005micromachined,banquy2009effect,liu2012uptake}. The behavior of the wrapping degree can be explained by the total energy of the system and a specific configuration in which the membrane wraps half of a rigid particle (Fig. \ref{fig6}-II-b). The adhesion to the particle for this specific case is denoted by $\omega=\omega_0 \approx 5.8 \, \varepsilon/{\sigma}^2$.
\begin{figure*}[ht!]
\centering
\includegraphics[width=1.6\columnwidth]{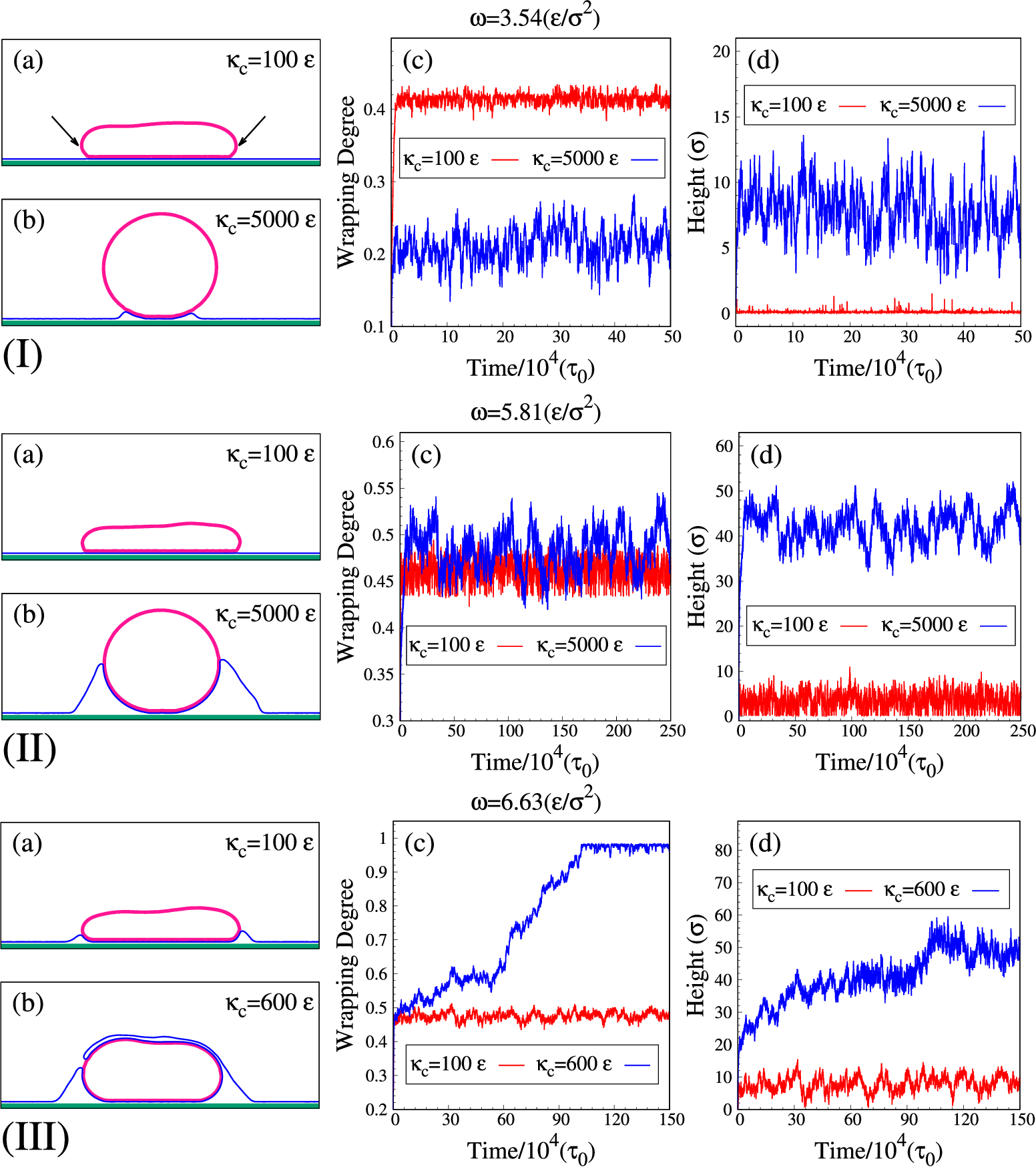}
\caption{ Comparing wrapping degree and the membrane protrusions height in soft and stiff particles for (I): $\omega=3.54\, \varepsilon/{\sigma}^2$; (II):$\omega=5.81\, \varepsilon/{\sigma}^2$; and (III):$\omega=6.63\, \varepsilon/{\sigma}^2$. In plots (I) and (II): panels (a) represent a typical configuration of a soft particle ($\kappa=100 \, \varepsilon$) which spread along the substrate. In this case, the membrane without any deformation adheres to nearly half of the particle (red line in panel (c)) but the height of membrane protrusions is negligible (red line in panel (d)). Panels (b) indicate that wrapping the membrane around a rigid particle ($\kappa=5000 \, \varepsilon$) needs deformations and detaching from the substrate. The blue lines in panel (c) and (d) denote the wrapping degree and height of the membrane for the rigid particle, respectively. Plot (II) demonstrates in $\omega=\omega_0 \approx 5.8 \, \varepsilon/{\sigma}^2$, rigid and soft particles more or less have same wrapping degree. However, in the rigid case, the height of the protrusions is remarkable and around the radius of the particle. In plot (III), because the adhesive interaction between the membrane and particle has been increased, in the soft case, membrane raise up a little bit around the target (panel (a) and red line in panel (c)) and less stiff particles can be completely wrapped (panel (b)).}
\label{fig6}
\end{figure*}
The total energy of the system with respect to the initial state, for which the membrane is fully adhered to the cytoskeleton, is written as,
\begin{equation}
\Delta E= -E_{ad}^{v}+E_{ad}^{s}+E_H+E_{b}^{v},
\end{equation}
where $E_{ad}^{v}$ and $E_{ad}^{s}$ are the adhesion energy of the membrane to the vesicle and the substrate, respectively. The Helfrich energy, $E_H$, represents the elastic energy of the membrane, while $E_{b}^{v}$ denotes the bending energy of the vesicle. For rigid particles, the last contribution vanishes. In the case of rigid targets, due to the adhesive energy between the membrane and the vesicle, the membrane detaches from the substrate, deforms, and wraps around the target.
 If $\omega < \omega_0$, and the particle is rigid, a competition between $E_{ad}^{v}$ and $E_{ad}^{s}+E_H$ leads to a wrapping degree smaller than $0.5$ (Fig. \ref{fig6}-I-b). On the other hand, soft vesicles have a lower bending rigidity, which allows them to spread along the substrate. As a result, the membrane can attach to about half of the vesicle without any deformation or detachment from the substrate. (Fig. \ref{fig6}-I-a). However, this induces two highly curved areas in the vesicle as shown in Fig. \ref{fig6}-I-(a) by the black arrows.  
Wrapping beyond these points requires a considerable amount of energy\cite{khosravani2, bahrami2013orientational}, and as a result, the membrane remains flat ($E_{ad}^{s}=E_H \approx 0$). If $\omega \approx \omega_0$, as our definition, the wrapping degree is $0.5$ for the rigid particle. In this criterion, a soft particle can also spread along the substrate and more or less has the same wrapping degree (Fig. \ref{fig6}-II). Wrapping degrees over 0.5 in rigid particles are achieved by increasing target adhesion energy, $\omega$, beyond $\omega_0$.
 However, the membrane protrusions cannot overcome the highly curved areas in the oblate shape of a soft vesicle, and the wrapping degree does not change ($\approx 0.5$). In our previous study, we derived a criterion for the minimum particle–membrane adhesion energy required for complete engulfment of an elliptical particle by a supported lipid membrane. 
For an elliptical particle with aspect ratio $D=a/b$, the critical adhesion energy for the transition from partial to full wrapped state is 
\begin{equation}
\omega_f=\frac{\kappa}{2a^2}\left[ \sqrt{\frac{4a^2}{k} \left( \Sigma+\omega_s \right) } + \frac{\left( \Sigma^2+D^2(\omega_s^2+2 \Sigma \omega_s) \right)^{3/2}}{D(\Sigma+\omega_s)^3} \right]^2,
\label{el_omega-f}
\end{equation}
where $2a$ and  $2b$ are the length of axes along the $x$ and $y$ directions, respectively. Accordingly, elliptical particles with larger aspect ratios require higher adhesion energy for full engulfment, analogous to soft particles spreading along the membrane.
\begin{figure}[b!]
\centering
\includegraphics[width=1\columnwidth]{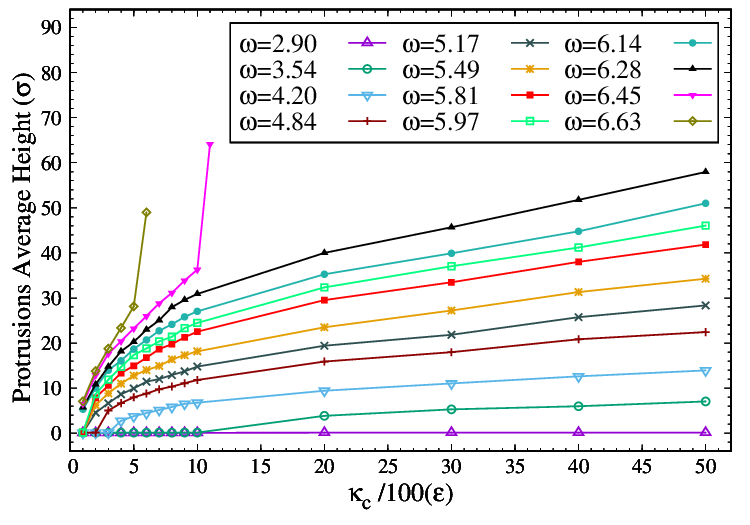}
\caption{The average of the membrane protrusions height as a function of $\kappa_c$ for different values $\omega$.}
\label{heights}
\end{figure}
In contrast with the wrapping degree, the height of the membrane protrusions is always an increasing function of $\kappa_c$ (Fig. \ref{heights}). This can be a sign of how macrophages selectively engulf aged red blood cells. According to experiments, aged red blood cells are more rigid than young ones\cite{marik1993effect,xu2018stiffness}, and it has been demonstrated that macrophages have a tendency for phagocyting the stiffer particles\cite{beningo2002fc,tao2005micromachined}. It is known that phagocytosis is a process in which cup shape protrusions of the membrane rise around the target and engulf it in a zipper-like manner. This process involves active forces from the cytoskeleton, which rely on a positive feedback mechanism possibly associated to membrane protrusions\cite{swanson2008shaping}. According to Fig. \ref{heights} the height of the protrusions is a decreasing function of the target flexibility. If the membrane cannot rise around the target the signaling process will be blocked and consequently, the process cannot be enhanced by active forces from the cytoskeleton. It should be mentioned that phagocytosis is a time-dependent process and if the engulfment to be postponed for a specific time, macrophages give up the process. Furthermore, it is worth mentioning that cortical forces have to experience a large direction change in the high curvature points created in soft particles, which can also decelerate the phagocytosis process. 
 On the other hand, viral particles like HIV and MLV enter the cells by such process as clathrin-mediated endocytosis, caveolae, and membrane fusion. It has been demonstrated that these viruses regulate their capsule stiffness for entry to cells as softer particles than they budding off their hosts\cite{kol2006mechanical,kol2007stiffness}. In clathrin-mediated endocytosis and caveolae, the height of the protrusions is not important, because in these processes the membrane usually deforms inward the cell to early pits evolve and enter the cells as early endosomes. A soft particle by spreading on the membrane increases the number of ligand-receptor interactions which leads to high rate endocytosis. As Fig. \ref{fig6}-(a) indicates a soft particle, even in small values of adhesion ($\omega \approx k_BT$), can spread along the membrane and internalize to cells. Finally, it is worth mentioning that a way for improving the efficiency of drug delivery vesicles can be regulation of their stiffness\cite{geng2007shape,masoud2012controlled}.\\
In this paper, we have studied the effect of particle stiffness on the cellular uptake by a molecular dynamics simulation. Our simulations show that the impact of the particle stiffness, $\kappa_c$, on the success of engulfment strongly depends on the values of membrane-target adhesion, $\omega$. It can be an explanation for different uptake rates reported in experiments with soft and rigid particles. In contrast with the amount of wrapping degree, the height of the membrane protrusions is an increasing function of $\kappa_c$. Because the extension of membrane protrusions is an essential step in the phagocytosis process, this result suggests that phagocytosis of rigid particles is more preferable than softer ones. On the other hand, soft particles lay on the membrane and it shows that endocytosis of soft particles can occur at a higher rate. In this study, we have performed a 2D simulation which is applicable in particles with high aspect ratio. However, we believe the behavior of the system is general and expandable to 3D problems or even cases with active cortical forces. Moreover, our simulations contain to parametric values, $\kappa_c$ and $\omega$, comparing these two parameters with experimental results enable us to tune the model for predicting the efficiency of cellular uptake in drug delivery applications.


\bibliography{achemso-demo}

\clearpage

\section{Supporting Information}
Membranes elastic energy is properly describable by the two dimensional integral of Helfrich,
\begin{equation}
E_H=\frac{1}{2}\kappa\int_a \left( 2H \right)^{2}da+\Sigma\int_a da,
\end{equation}
where $H$ is the mean curvature of the membrane. Long particles have the symmetry in their long axis which helps writing the elastic energy per unit length of the particle. Consequently, the 1D Helfrich energy can be discretized as a chain of beads which connected to each other by a harmonic spring potential,
\begin{equation}\label{spring_potential}
E_{spring}=\frac{1}{2} \Lambda \sum_{i=1}^{N-1} \left[ {d\left(i\right)-d_0}\right]^{2}.
\end{equation}
$\Lambda=5000 \, \varepsilon/\sigma^2$ is the stiffness of the springs per unit length of the particle, $d\left(i\right)$ is the bond length, and $d_0$ is the equilibrium bond length. The bending energy of the membrane preserves as the following potential between each 3 connected beads (Fig. \ref{fig1}-(b)),
\begin{equation}\label{bending}
E_B=\frac{\kappa}{d_0} \sum_{i=1}^{N-2} \left[1-\cos \left(\theta \left(i,i+1\right) - \theta_0  \right)  \right], 
\end{equation}
where $\kappa$ is the bending rigidity of the membrane, and $\theta(i,i+1)$ represents the angle between neighboring springs which their equilibrium value is $\theta_0$. Furthermore Adding a lateral force at the edge of the membrane reproduces the effect of the membrane tension $\Sigma=1.5 \, \varepsilon/\sigma^2$. Here, the membrane is constructed by $1000$ monomers, is allowed to move only in $x-y$ plane, and the excluded volume interactions between the membrane beads are implemented using the Weeks-Chandler-Andersen potential
\begin{align}\label{WCA}
V_{WCA}\left(r_{ij}\right)=\left\lbrace 
\begin{array}{cc}
4\varepsilon \left[ \left( \frac{\sigma}{r_{ij}}\right) ^{12}-\left( \frac{\sigma}{r_{ij}}\right) ^{6}+\frac{1}{4} \right],  & r_{ij}\leq r_c\\
0, & r_{ij}> r_c,
\end{array} \right.
\end{align}
where $r_c =2^{1/6} \, \sigma$, $r_{ij}$ is the distance between the $i$th and $j$th beads and $\varepsilon$ and $\sigma$ are the unit energy and length scale of the simulation, respectively. The diameter of the membrane's monomers is $R=d_0 \doteq 1\sigma$. The cytoskeleton which laid underneath the membrane is also implemented as a collection of immobile beads. The distance between its constructive monomers was considered $0.5 \, \sigma$ to simplify the reptation of the membrane on the cytoskeleton.

The initially cylindrical vesicle is assembled from $630$ monomers which connected to each other by potentials (\ref{spring_potential}) and (\ref{bending}) and they are not allowed to intersect each other using the WCA potential, Eq. (\ref{WCA}). In the vesicle structure, $d_0 \approx 0.5 \, \sigma$, $\Lambda=1000 \, \varepsilon/\sigma^2$, $\kappa=50 \, \varepsilon$, and $\theta_0=(\pi -2\pi/630)$. The particle is placed in the center and on top of the membrane and its initial radius is $a= 50 \, \sigma$. The elasticity of the vesicle is modulated with a bending potential, Eq. (\ref{bending}), which only applied to every 10th monomers of vesicle (totally 63 monomers of the vesicle that some of them is represented in Fig. \ref{fig1}(a) by black points). In this case, $\theta_0=\theta_c=((\pi -2\pi/63))$, and $\kappa=\kappa_c$.\\
Both the membrane-cytoskeleton and the membrane-particle interactions are modeled with the following potential:
\begin{eqnarray}
V\left(r_{ij}\right) = 
  \begin{cases}
 4\lambda_{k} \left[ \left( \frac{\sigma}{r_{ij}}\right) ^{12}-\left( \frac{\sigma}{r_{ij}}\right) ^{6} \right],  &r_{ij} < r_{c} \\
-\lambda_{k} \cos^{2}\left[\frac{\pi}{2\zeta}\left(r_{ij}-r_{c}\right)\right] , &r_c \leq r_{ij} \leq r_{c}+\zeta \\
0,  &r_{ij} > r_{c}+\zeta,
  \end{cases} \nonumber
  \label{eq:dt}
\end{eqnarray}
where $\lambda_k$, in the unit of the energy, denotes the strength of the ligand-receptor interactions ($k=1$ corresponds to the membrane-vesicle, and $k=2$ corresponds to the membrane-cytoskeleton), and $\zeta=0.5 \, \sigma$. The values of the average adhesion energy per unit length $\sigma$ between the membrane and cytoskeleton ($\omega_s$) and between the membrane and the particle ($\omega$) can be tuned by varying $\lambda_1$ and $\lambda_2$. 

Our Molecular Dynamics (MD) simulations were performed at the constant temperature $T=1.0 \, \varepsilon/k_B$, with the Langevin thermostat, and using ESPResSo. The time step in the Verlet algorithm and the damping constant in the Langevin thermostat were set $\delta t=0.01 \tau_0$ and $\Gamma=\tau^{-1}_0$, respectively, which $\tau_0=\sqrt{\frac{m\sigma^{2}}{\varepsilon}}$ is the MD time scale and $m$ is the monomer mass.

\end{document}